# Raman and optical characterization of multilayer turbostratic graphene grown via chemical vapor deposition

Daniel R. Lenski[†], Michael S. Fuhrer[*]

*Department of Physics and Center for Nanophysics and Advanced Materials, University of Maryland,*

*College Park, Maryland 20742-4111, USA*

We synthesize large-area graphene via atmospheric-pressure (AP) chemical vapor deposition (CVD) on copper, and transfer to $SiO_2$ wafers. In contrast to low-pressure (LP) CVD on copper, optical contrast and atomic force microscopy measurements show AP-CVD graphene contains significant multi-layer areas. Raman spectroscopy always shows a single Lorentzian *2D* peak, however systematic differences are observed in the *2D* peak energy, width, and intensity for single- and multi-layer regions. We conclude that graphene multi-layers grown by AP-CVD on Cu are rotationally disordered.

† Present address: Intel Corporation, 5200 NE Elam Young Parkway, Hillsboro, OR 97124
* Electronic mail: mfuhrer@umd.edu

## I. Introduction

With its high field-effect mobility and optical transparency, graphene is a promising material for future electronics applications, including both complementary integrated circuits and optically transparent electrodes for displays and sensors.[1]  Mechanical exfoliation of graphite[2] has produced the highest-performing samples, but the manual effort, unreliable results, and small areas hinder practical applications of graphene.[3,4]  Other synthesis methods include epitaxial growth on SiC and chemical vapor deposition on single-crystalline metals, both of which require costly substrates.  Chemical vapor deposition (CVD) on polycrystalline nickel or copper,[5-9] in thin films or foils, can produce wafer-scale graphene at low cost.[5,9]

## II. Experimental methods

Previous reports of low-pressure CVD on copper demonstrate that graphene forms by surface adsorption, and is thus self-limited to a single graphene monolayer, which passivates the metal surface.[3]  However, recently the growth of large areas of AB-stacked bilayer graphene was reported in low-pressure CVD on copper.[10]  Here we report on the growth and characterization of graphene on copper substrates by CVD at atmospheric pressure.[5]  We characterize our graphene films by optical contrast and Raman spectroscopy.[11-13]  We identify regions of multi-layer graphene as evidenced by optical contrast.  Raman spectroscopy of these regions shows a single-Lorentzian *2D* peak indicative of rotationally-disordered graphene, and the *2D* peak shows systematic differences in position, width, and intensity from regions of single-layer graphene.

We adopt a growth procedure similar to that of Lee et al.[5] to grow graphene on Cu foils of 99.8% purity and 25 μm thickness (Alfa Aesar #13382).  Foils are cleaned by sonication in acetone, then rinsed in methanol followed by isopropanol, and then loaded into a 1 inch quartz tube furnace.  Foils are heated to approximately 1050°C under atmospheric pressure in flowing Ar:$H_2$=1000:50

SCCM and held for 30 minutes before graphene growth in $CH_4$:$H_2$:Ar=50:15:1000 SCCM, then cooled to room temperature in Ar:$H_2$=1000:50 SCCM. Growth times were varied from 30-300 s, with no apparent variation in the resulting graphene films. Cooling rate was also varied from <1 °C/s to >10 °C/s, and again no variations were found, contrasting with reports of graphene grown on Ni via atmospheric-pressure CVD.[14,15] Variation of temperature does have a substantial effect: samples grown at lower temperatures showed more defects (larger Raman *D* band intensity) and substantial deposits of opaque material with a Raman spectrum suggestive of amorphous carbon, with no signature of graphene observed in Raman at growth temperatures below approximately 950°C..

Following growth, graphene is transferred from the Cu foils using a procedure substantially similar to that of Reina et al.[6] Poly[methyl methacrylate] (PMMA) is spun-cast on the foils, and dried briefly at 150°C. Foils are then immersed in Transene APS-100 etchant and heated to 60-80°C. The copper is entirely etched within 30 min and the nearly transparent films of graphene and PMMA float to the surface. Films are rinsed in DI water, then transferred onto chips of nominal 300 nm thermally-grown $SiO_2$ on Si. The chips are dried at 60°C, the PMMA is dissolved in acetone, and finally the chips are rinsed in isopropanol.

### III. Raman and optical measurements

Figure 1(a) shows a representative sample of CVD graphene in an optical microscope [Olympus STM6 with 100×, 0.9 numerical aperture objective]. A void is intentionally introduced into the graphene by tearing during transfer. Non-uniform optical contrast over the graphene is immediately apparent: while a majority of the graphene shows fairly uniform contrast against the bare $SiO_2$ (lower right), there are regions of greater contrast, which appear to occur in discontinuous parallel bands of roughly 3-10 μm width that are apparent on a larger scale (see Supplemental Material for larger images). These are suggestive of the distribution of parallel polishing marks on the as-supplied copper foils. Figure 1(b) shows histograms of the reflected intensity from several regions of Figure 1(a), compared with a reference sample containing two flakes of mechanically-exfoliated graphene (one of which is confirmed by Raman spectroscopy as a Bernal-stacked bilayer[11]). Figure 1(b) clearly shows multiple peaks in the intensity histograms for both CVD and mechanically-exfoliated samples, suggesting discrete thicknesses of graphene; we discuss this more quantitatively below.

Blake et al.[16] calculated the relationship between illumination wavelength, oxide thickness, and the contrast of monolayer graphene on $SiO_2$ (defined as $1-R_g/R_s$, where $R_g$ and $R_s$ are the reflectance of the graphene and the substrate, respectively). While Blake et al. predicted a nearly constant contrast per graphene layer, assuming a refractive equal to that of bulk graphite, other reports show considerable sample-to-sample variation, thickness-dependence, and dispersion of the index of refraction of graphene, $n_g$.[17-19] Using $n_g = 2.0 - 1.1\,j$ from Ref. 17 we find that the observed contrast of our mechanically-exfoliated graphene samples correspond well to theoretical predictions for 306 nm $SiO_2$ (the vendor-specified mean thickness of the sample shown in the inset of Figure 1(a)); dashed vertical lines show calculated 1-4 layer contrast in Figure 1(b). In particular, the peak for the bilayer exfoliated flake at 16% contrast matches the theoretical calculation. The regions of CVD graphene in Figure 1(a) show peaks at 12-13% contrast, probably corresponding to monolayer graphene. Although this contrast does not match the prediction for monolayer graphene on 309 nm $SiO_2$ (again, the vendor-specified mean thickness) using $n_g$ from Ref. 17, it is within the range predicted by experimentally-measured values for graphene's index of refraction.[19] Regions F and G show secondary peaks around 19% contrast, which is also within the plausible range for bilayer graphene. Similar multi-modal contrast histograms are observed for other samples (see Supplemental Material).

Figure 2 shows micro-Raman spectra of CVD graphene at representative low-contrast and high-contrast points, with Lorentzian fits to the observed peaks. We used a Horiba J-Y Raman microscope with laser wavelengths of 514 nm and 633 nm, at low power to reduce sample heating. The *G* and *2D* peaks are prominent Raman features that can be used to distinguish 1-5 layers of AB-stacked graphene.[11] In these and other Raman spectra of CVD graphene (summarized in the Supplemental Material), we observe significant variation in *G* peak position and width, which may be attributed to electronic doping of the graphene by the substrates.[11,20] Surprisingly, all of our Raman spectra show narrow, symmetrical *2D* peaks, even in areas where optical contrast indicates multilayer thickness. On our CVD graphene we have *never* observed the broad, multi-peaked *2D* band of AB-stacked multilayer graphene,[5,6,13] and which we readily reproduce on a bilayer sample of mechanically exfoliated graphene (topmost spectrum in Figure 2). (See Supplemental Material for more Raman spectra.)

While the *2D* band of our CVD graphene is always single-peaked, it varies in other ways between regions of lower and higher optical contrast. Figure 3 shows three features of the Lorentzian fits to the *2D* peak measured at several points on CVD graphene with both higher and lower optical contrast. Figure 3(a) shows a blue-shift of about 8 cm$^{-1}$ from the *2D* peak position of lower-contrast CVD graphene to that of higher-contrast CVD graphene, while Figure 3(b) shows a broadening of the *2D* peak in higher-contrast graphene, and Figure 3(c) shows a reduction in the relative intensity of the *2D* peak of the higher-contrast graphene. These effects are apparent in Raman spectra measured using both 514 nm and 633 nm laser excitation.

## IV. Discussion

Poncharal et al.[21] showed that the Raman *2D* band of misoriented bilayers of exfoliated graphene resembles that of monolayer graphene but blue-shifted relative to monolayer graphene. Faugeras et al. also observed this blue-shift in few-layer graphene grown on the $(000\bar{1})$ face of SiC,[22] while Pimenta et al. observed it along with increased *2D* peak width in many-layer turbostratic graphite.[23] Based on the evidence of steplike variation in optical contrast and systematic deviations in Raman spectra between high- and low-contrast regions, we conclude that the high-contrast regions of our CVD graphene are misoriented or turbostratic multilayer graphene. This observation contrasts with the recent claim of growth of large areas of AB-stacked bilayer graphene on copper by low-pressure CVD[10] (a different process than that used here). Interestingly, the Raman *2D* mode for bilayer graphene in Ref. 10 also appears broader and more symmetric than that of exfoliated AB-stacked graphene (similar to our observations), suggesting the possibility of rotational disorder in those samples as well.

We also used a Veeco DI-5000 atomic force microscope (AFM) in tapping mode and ambient conditions to obtain topography images of flakes of CVD graphene transferred to $SiO_2$. It was difficult to find large, flat areas around the edges of the graphene which were not contaminated with particles (likely PMMA from the transfer process), but a few were located. Figure 4(a) is a tapping-mode AFM topography image takene near the edge of a piece of CVD graphene transferred to nominal 300 nm $SiO_2$. A step from the $SiO_2$ substrate to the graphene flake is visible, as shown in Figure 4(b). After the initial step from the $SiO_2$ substrate, there is another step, indicating one or more additional monolayers of graphene.

We consider alternative explanations for the combination of monolayer-like Raman spectra and optical contrast and topography corresponding to those of multilayer graphene. After the transfer process, it is possible that a layer of water remains under the graphene in some areas, or a thin layer of PMMA residue may remain on top. Alternatively, the graphene may not adhere closely to the $SiO_2$, leaving an air gap between. A third possibility is that some thin layer of PMMA residue remains on top of the graphene film.

We have studied the effect of additional dielectric layers by extending the transfer-matrix method used Blake et al. to a 4-layer structure. We calculate the contrast of graphene layers on $SiO_2$ with up to 4 nm of water (*n*=1.333*)* or air (*n*=1.0) between the graphene and the substrate, or up to 4 nm of PMMA (*n*=1.49 at *λ*=600 nm) on top the contrast of graphene layers on $SiO_2$ with up to 4 nm of water ($n$=1.333) or air ($n$=1.0) between the graphene and the substrate, or up to 4 nm of PMMA ($n$=1.49 at $\lambda$=600 nm) on top (see Supplemental Material), and in general find that any of these impurities could slightly *reduce* the contrast of graphene on 309 nm $SiO_2$ (the thickness on which our CVD graphene was deposited), thus causing a multilayer sample to look like a monolayer sample, but that they cannot *increase* its contrast. Furthermore, the contrast shift introduced by these impurities is nearly independent of the graphene thickness, and so cannot significantly affect the spacing of contrast peaks for different graphene layer numbers. We therefore rule out contamination as a possible explanation for our observed multimodal optical contrast histograms.

In conclusion, we have grown graphene on Cu foils via atmospheric-pressure chemical vapor deposition. We observe variation in the optical contrast of graphene films transferred to $SiO_2$ wafers, and peaks in the contrast histograms are consistent with the presence of multilayered graphenes. Raman spectroscopy shows single-peaked *2D* bands on all samples of CVD graphene, but with systematic variations in peak position, width, and relative intensity according to layer number (as indicated by optical contrast), indicating that all multilayer regions consist of misoriented or turbostratic layers. Our results highlight the difficulties of conclusively measuring CVD graphene thickness: impurity layers can reduce the contrast of multilayer graphene on $SiO_2$, while misorientation can mimic the single Raman *2D* peak of single-layer graphene. The ratio of *2D* and *G* peak amplitude, often used to estimate graphene thickness, is also unreliable due to non-uniform adhesion of graphene to its substrate[13] and to doping.[20] While we do not fully understand the mechanism by which multilayer graphene forms on copper at atmospheric pressure, its distribution suggests a relationship to the topography of the copper substrates used.

The authors would like to thank Jianhao Chen for supplying samples of mechanically-exfoliated graphene used for comparison in this work. This research was supported by the Laboratory for

## Figure 1

(a) Contrast-stretched optical micrograph of CVD graphene transferred to 309 nm SiO$_2$ substrate under $\lambda$=600 nm illumination. Inset shows a reference sample of two mechanically exfoliated graphene flakes on a background of 306 nm SiO$_2$. (b) Histograms of reflected light intensity from the marked regions of (a) normalized to equal intensity for the SiO$_2$ substrate. Dashed lines show theoretical peak positions for 0-4 layer graphene for the corresponding substrate thickness (*see main text*).

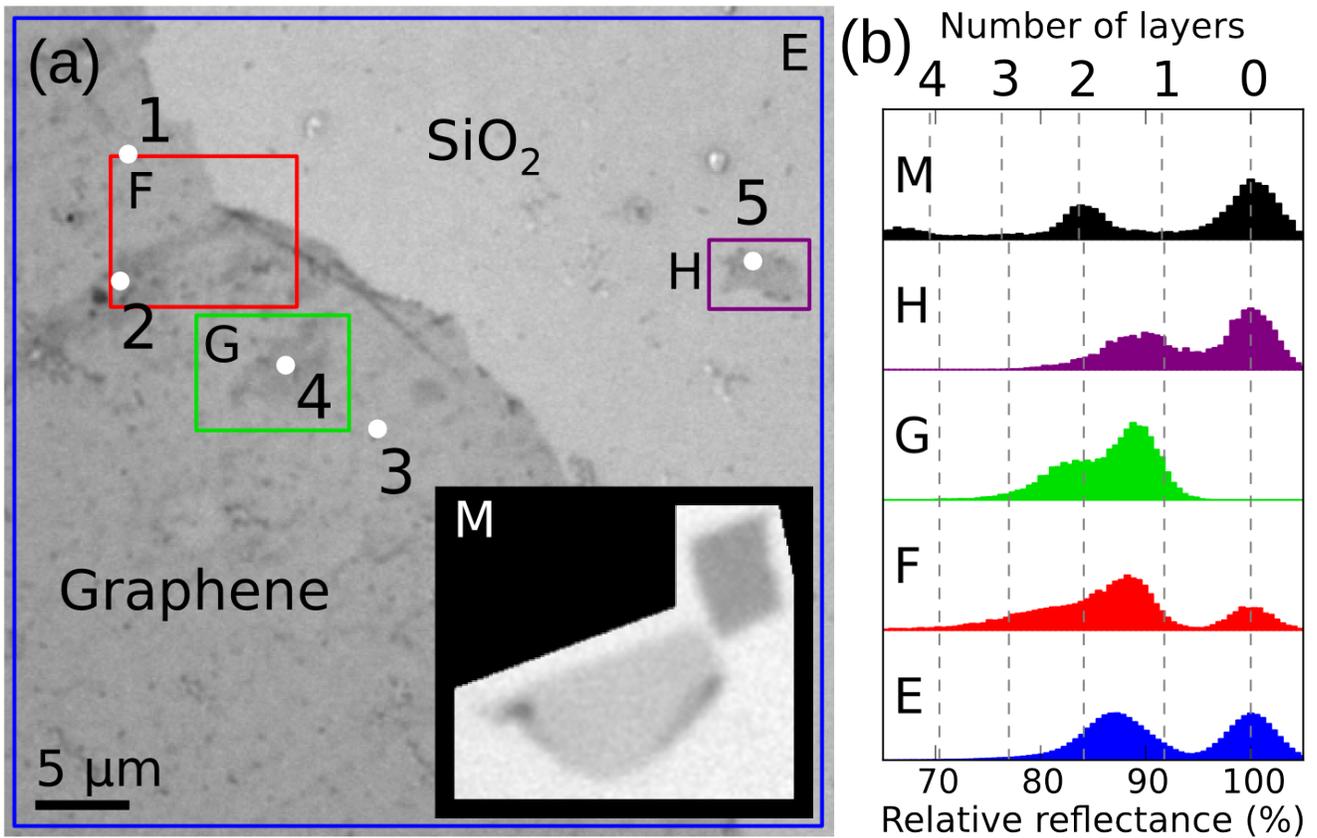

**Figure 2**

Representative micro-Raman spectra of CVD graphene at points 1 and 2 in Figure 1 with low (~12-13%, point 1) and high (~19%, point 2) contrast, along with comparative spectra of a mechanically exfoliated bilayer graphene (MEBLG) sample, all normalized to equal *G* peak height. Solid curves show Lorentzian fitting peaks, while dashed lines show Raman background.

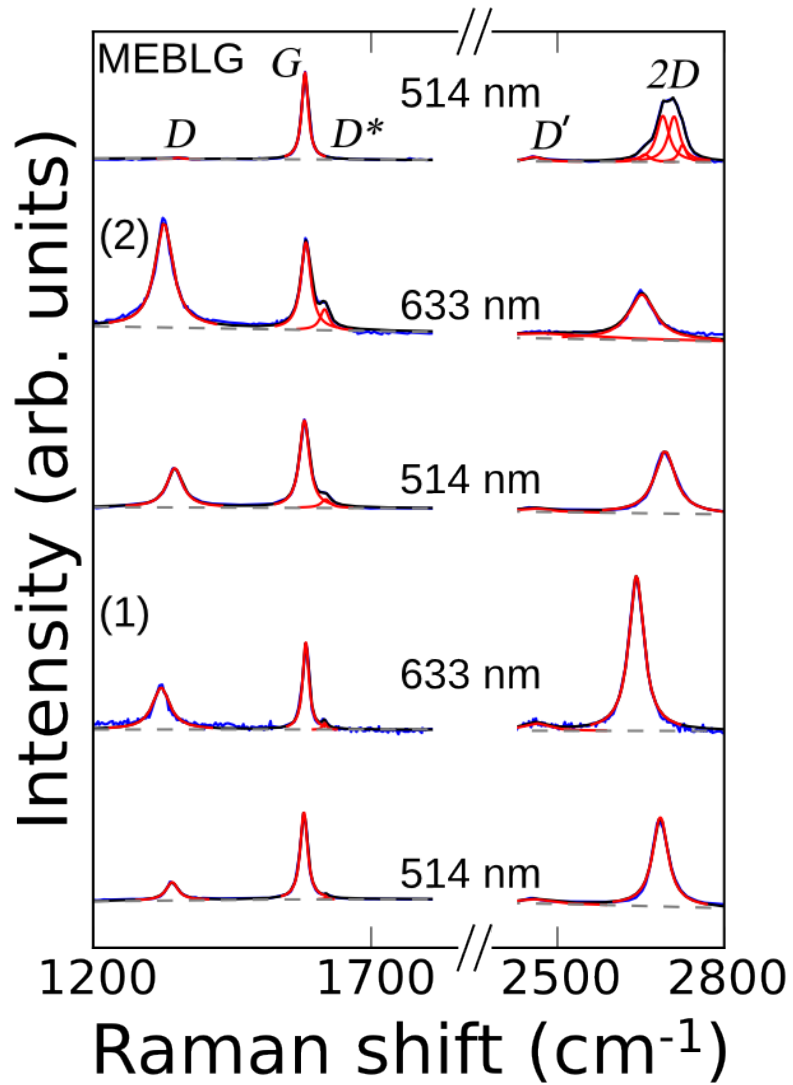

**Figure 3**

(a) *2D* peak positions, (b) *2D* peak widths, and (c) relative intensity of *2D* peaks for 514 nm (left) and 633 nm (right) Raman spectra on the CVD graphene sample shown in Figure 1 (points 1-5) and an additional sample (points 6-9; see Supplemental Material). Blue circles indicate points of lower (~12-13%) optical contrast, while red triangles indicate points of higher (~19%) contrast. Shaded bands indicate $\langle x \rangle \pm \sigma_x$ ranges for each peak parameter, grouped by contrast. (Note: 633 nm Raman spectra were not measured for point 7.)

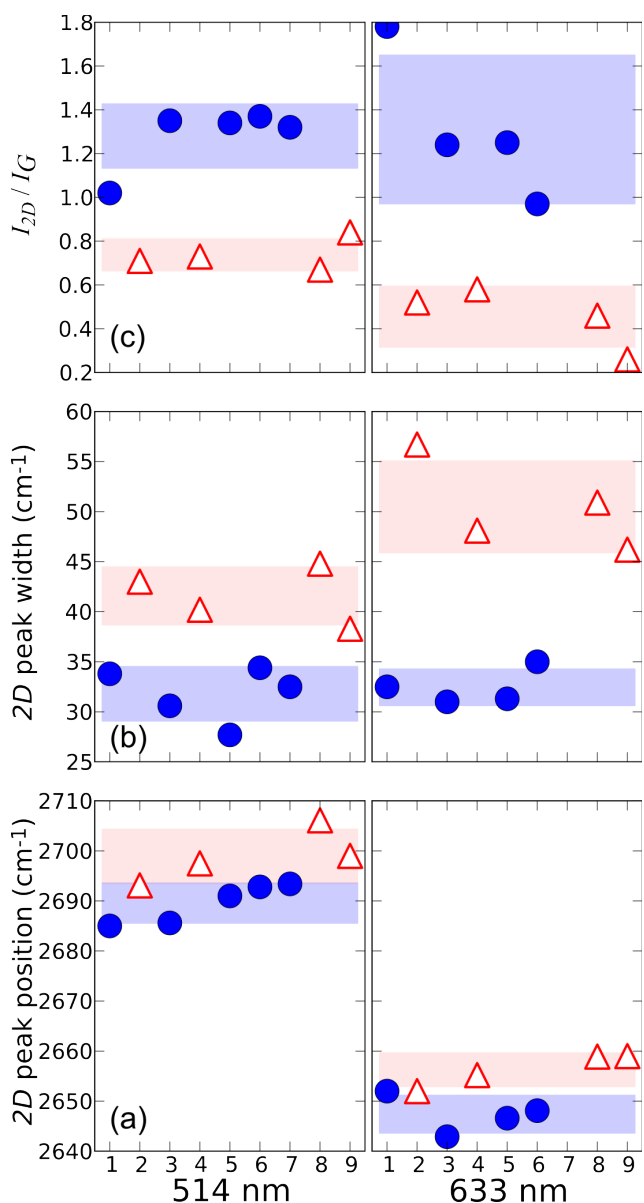

**Figure 4**

(a) Tapping-mode ambient AFM topography image around the edge of a piece of CVD graphene transferred from copper to nominal 300 nm $SiO_2$.  (b) Line profile of sample height (in nm) along the length of the dashed box in (a), averaged across its width.  A step from the $SiO_2$ substrate to the graphene is visible, followed by a second step indicating one or more monolayers of graphene on top of the first.  The arrow in (a) indicates another such step visible nearby on this sample.  Crosses in (a) and dashed vertical lines in (b) indicate edges of the first step from $SiO_2$ to graphene.

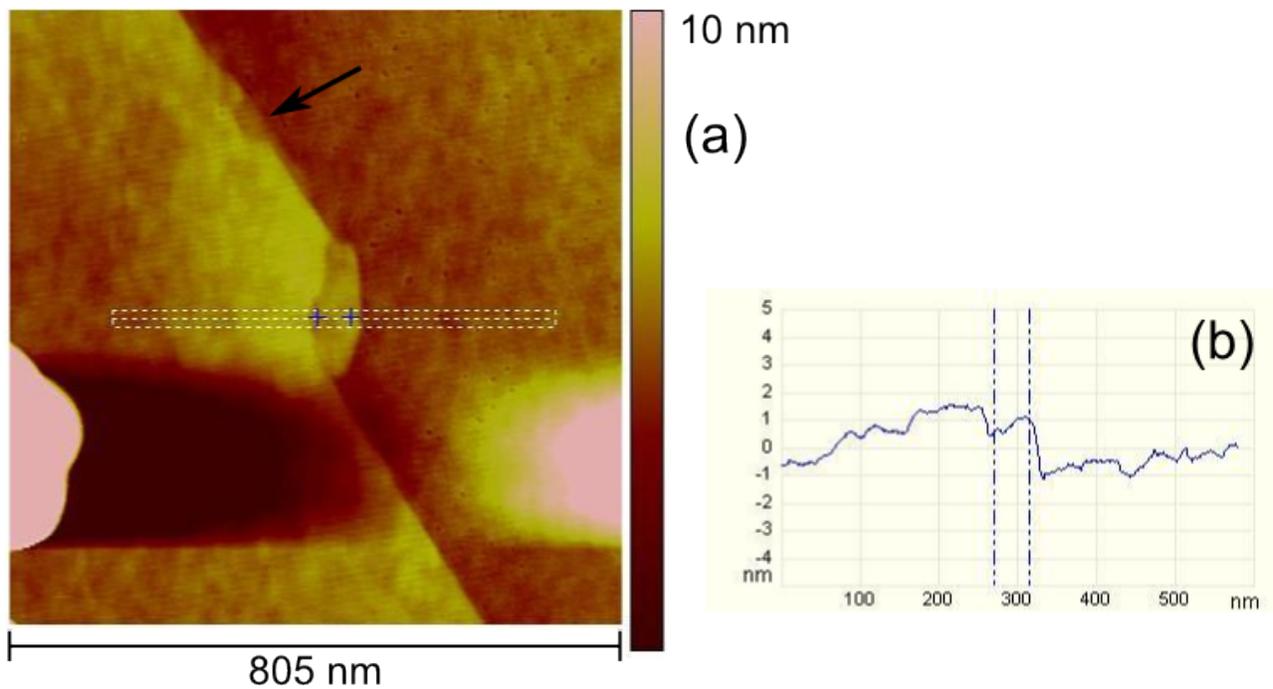

# Supplemental Material

## *I. Optical and Raman data from an additional sample of CVD graphene on SiO₂*

Additional samples of CVD graphene were produced in the same fashion as the sample shown in Figure 1 of the main text, and showed similar optical and Raman properties. Figure S1 and Figure S2 show properties of one such sample, made in a completely separate batch, weeks apart from that shown in Figure 1).

Figure S1 shows optical contrast of this sample, in the same fashion as Figure 1, and compares it with the optical contrast of the same mechanically-exfoliated graphene flakes as shown in Figure 1.

Figure S2 shows representative Raman spectra from this sample, in the same fashion as Figure 2, and compares them with the Raman spectrum of a *different* mechanically-exfoliated graphene bilayer. This bilayer is notable because its Raman spectrum shows a substantial *D* peak, but otherwise conforms to the Bernal-stacked bilayer spectra of Ferrari et al.[11]

**Figure S1**

(a) Contrast-stretched optical micrograph of CVD graphene transferred to 309 nm SiO$_2$ substrate under $\lambda$=600 nm illumination.  Inset shows a reference sample of two mechanically exfoliated graphene flakes on a background of 306 nm SiO$_2$. (b) Histograms of reflected light intensity from the marked regions of (a) normalized to equal intensity for the SiO$_2$ substrate.

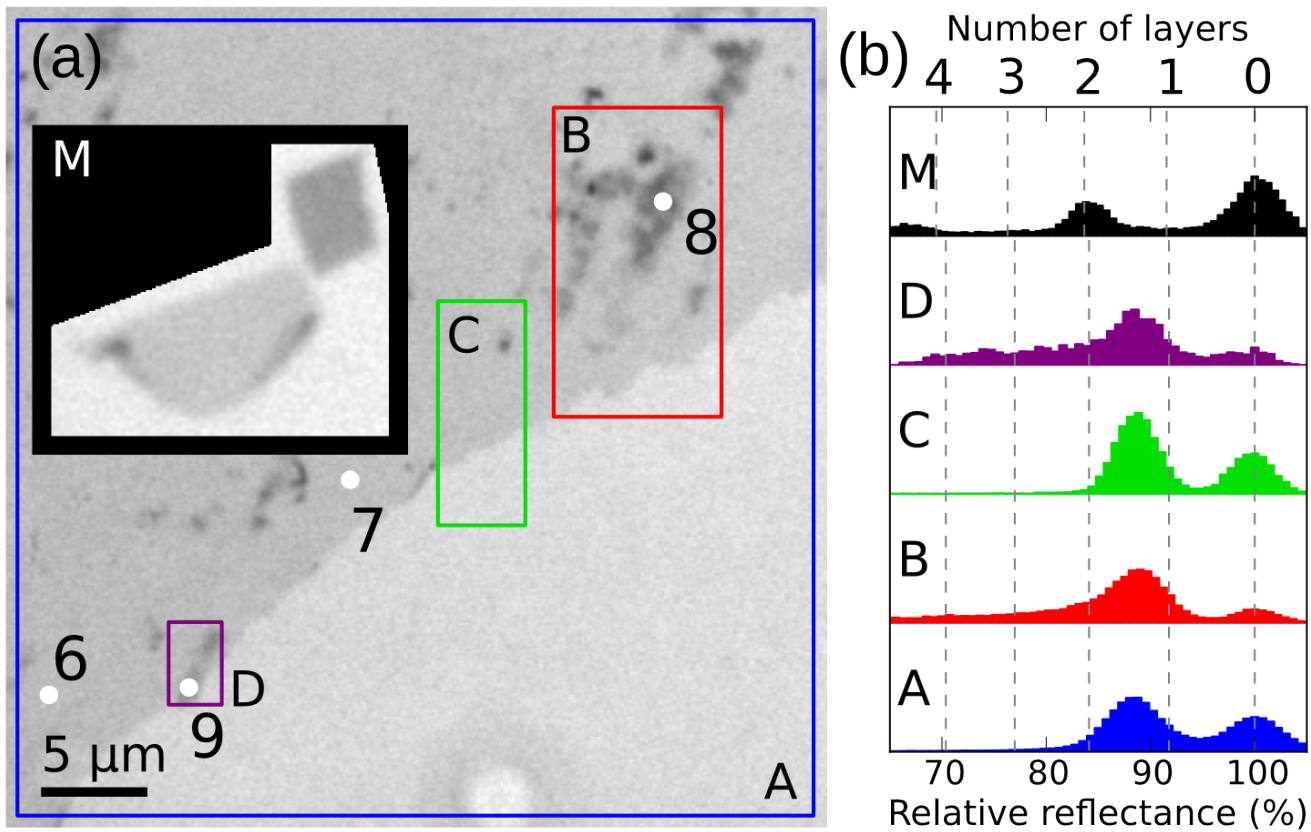

**Figure S2**

Representative micro-Raman spectra of CVD graphene at points 6 and 8 in Figure S1 with low (~12-13%, point 6) and high (~19%, point 8) contrast, along with comparative spectra of a mechanically exfoliated bilayer graphene sample, all normalized to equal *G* peak height. Solid curves show Lorentzian fitting peaks, while dashed lines show Raman background.

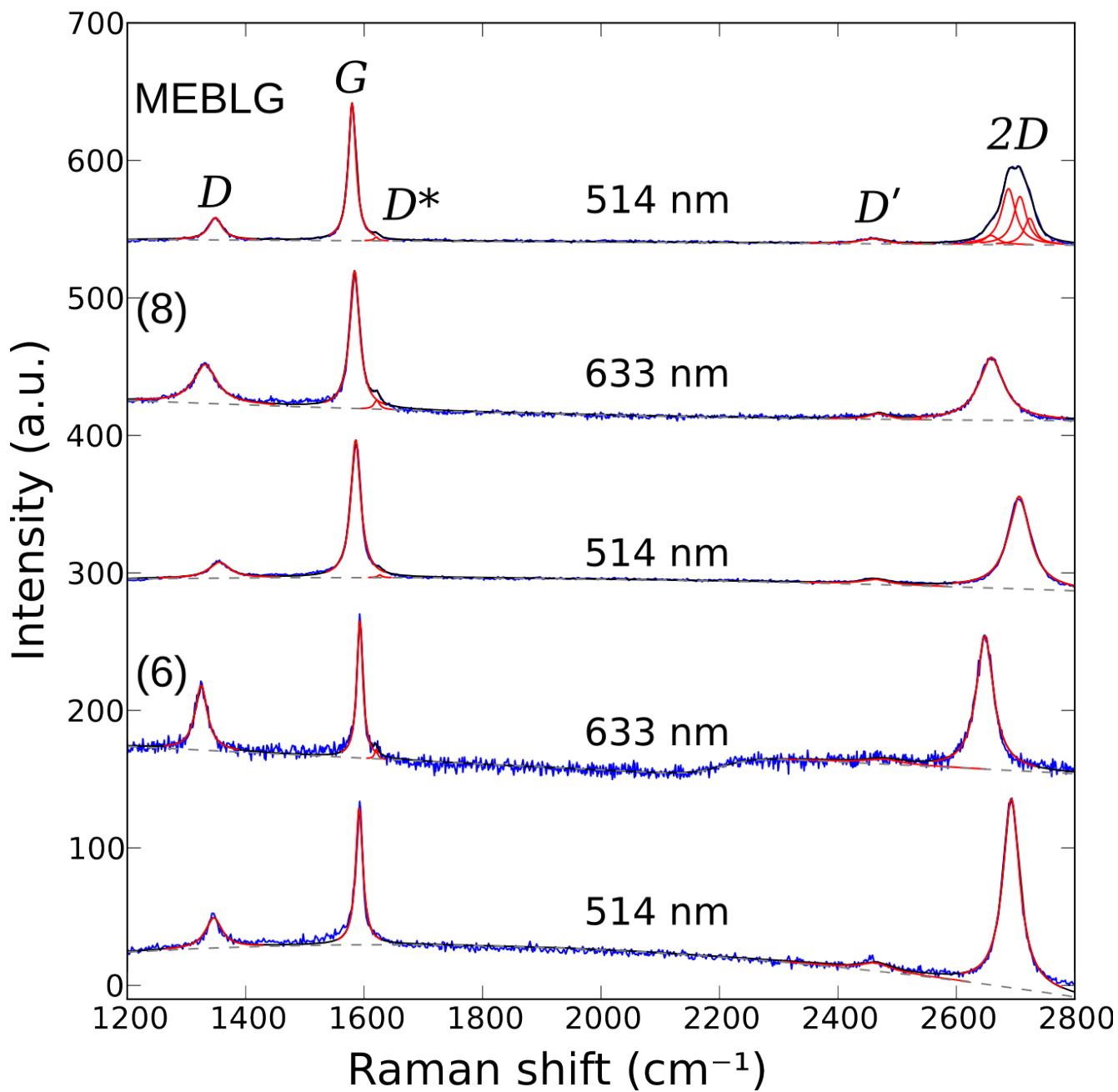

## II. Distribution of high-contrast graphene in parallel bands

As discussed in the main text, we found regions of higher optical contrast on our CVD-grown graphene to be predominantly distributed in parallel bands. Figure S3 shows larger-scale images of the CVD graphene samples shown in Figure 1(a) and Figure S1(a), making this distribution more visually apparent.

**Figure S3**

Wider contrast-stretched optical micrographs of the regions of CVD graphene shown in in Figure 1(a)

and Figure S1(a), under λ=600 nm illumination.  The insets of the images indicate the boundaries of the sub-regions depicted in Figure 1(a) and Figure S1(a).

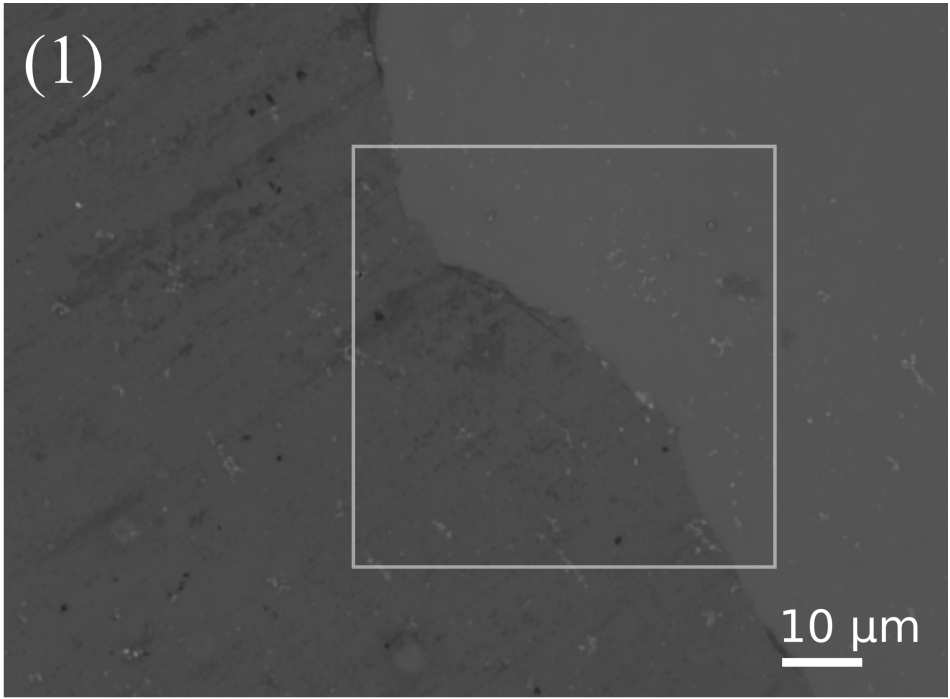

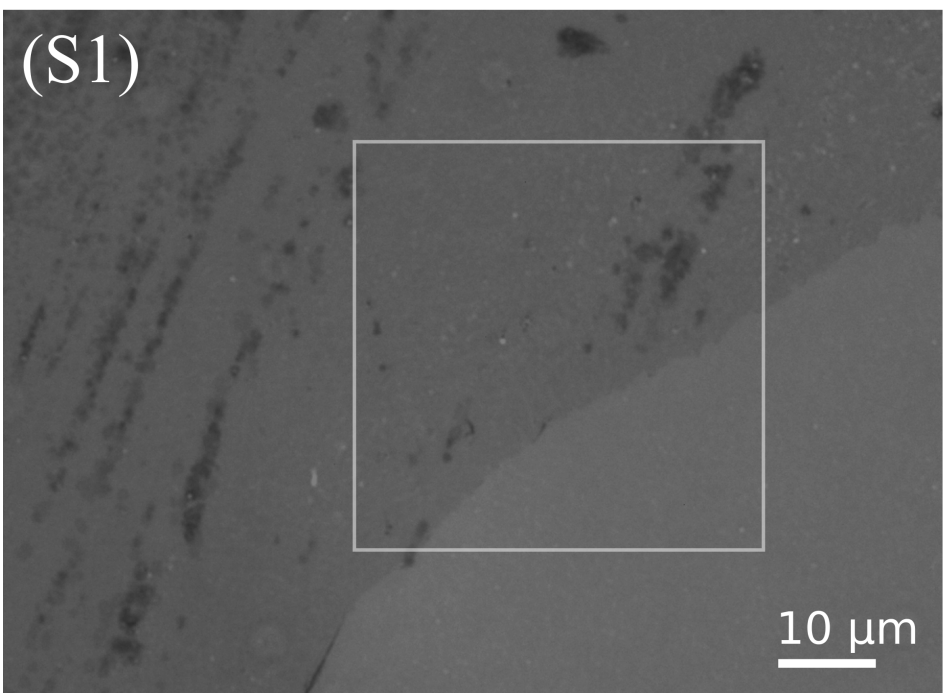

### III. Effect of impurities on graphene contrast

As discussed in the main text, we studied the effect of additional dielectric layers by extending the transfer-matrix method used by Blake et al.[16] to a 4-layer structure. We calculated the contrast of graphene layers on $SiO_2$ with up to 4 nm of water or air between the graphene and the substrate, or up to 4 nm of PMMA on top.

We extended the model of Blake et al.[16] from a 3-layer structure (graphene on $SiO_2$ on Si) to a 4-layer structure including an impurity layer either (a) below or (b) above the graphene:

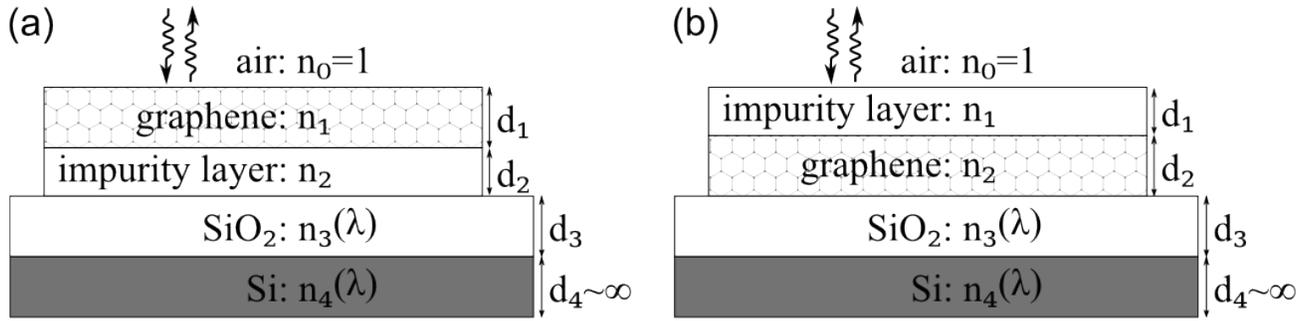

Using the transfer-matrix method, we can calculate the intensity reflectance coefficient of these four-layer structures, assuming normal incidence of light. The formula is substantially lengthier than that for a 3-layer structure, but an exact analytical expression can be derived:

$$
\begin{aligned}
A = {} & r_{01} + r_{12} e^{2j\Phi_1} + r_{23} e^{2j(\Phi_1+\Phi_2)} + r_{34} e^{2j(\Phi_1+\Phi_2+\Phi_3)} + r_{12} r_{23} r_{34} e^{2j(\Phi_1+\Phi_3)} \\
& + r_{01} r_{12} r_{23} e^{2j\Phi_2} + r_{01} r_{12} r_{34} e^{2j(\Phi_2+\Phi_3)} + r_{01} r_{23} r_{34} e^{2j\Phi_3} \\
B = {} & 1 + r_{01} r_{12} e^{2j\Phi_1} + r_{01} r_{23} e^{2j(\Phi_1+\Phi_2)} + r_{01} r_{34} e^{2j(\Phi_1+\Phi_2+\Phi_3)} \\
& + r_{01} r_{12} r_{23} r_{34} e^{2j(\Phi_1+\Phi_3)} + r_{12} r_{23} e^{2j\Phi_2} + r_{12} r_{34} e^{2j(\Phi_2+\Phi_3)} + r_{23} r_{34} e^{2j\Phi_3} \\
R = {} & \left| A/B \right|^2
\end{aligned}
\tag{S1}
$$

where $r_j = (n_j - n_{j+1})/(n_j + n_{j+1})$ is the coefficient of reflection at the interface below layer $j$, $\Phi_j = -2 k_j d_j = -4\pi n_j d_j/\lambda$ is the phase shift of a round-trip optical path through layer $j$, $n_j$ is the index of refraction of layer $j$, and $d_j$ is the thickness of layer $j$. Contrast is defined, as in Blake et al.,[16] by the relative intensity of reflected light in the presence and absence of the graphene and impurity layers:

$$C = \frac{I_r(n_1=1, n_2=1) - I(n_1, n_2)}{I_r(n_1=1, n_2=1)} \quad \text{(S2)}$$

Figure S4 shows the calculated contrast of 1-4 layers of graphene on 309 nm $SiO_2$, in the presence of impurity layers, and assuming $n_g = 2.0 - 1.1\,j$ as the index of refraction of graphene.[17]

**Figure S4**

Theoretical calculations of the contrast of 1-4 layer graphene (using Equations S1 and S2) on 309 nm SiO$_2$ under 600 nm illumination, with an air or water layer underneath or a PMMA layer above. The curves show the contrast of the graphene *without* the impurity layer, assuming $n_g = 2.0 - 1.1\,j$ as the index of refraction of graphene. The error bars show the range of contrasts with 0-4 nm of the impurity layer. Curves are slightly offset horizontally for clarity.

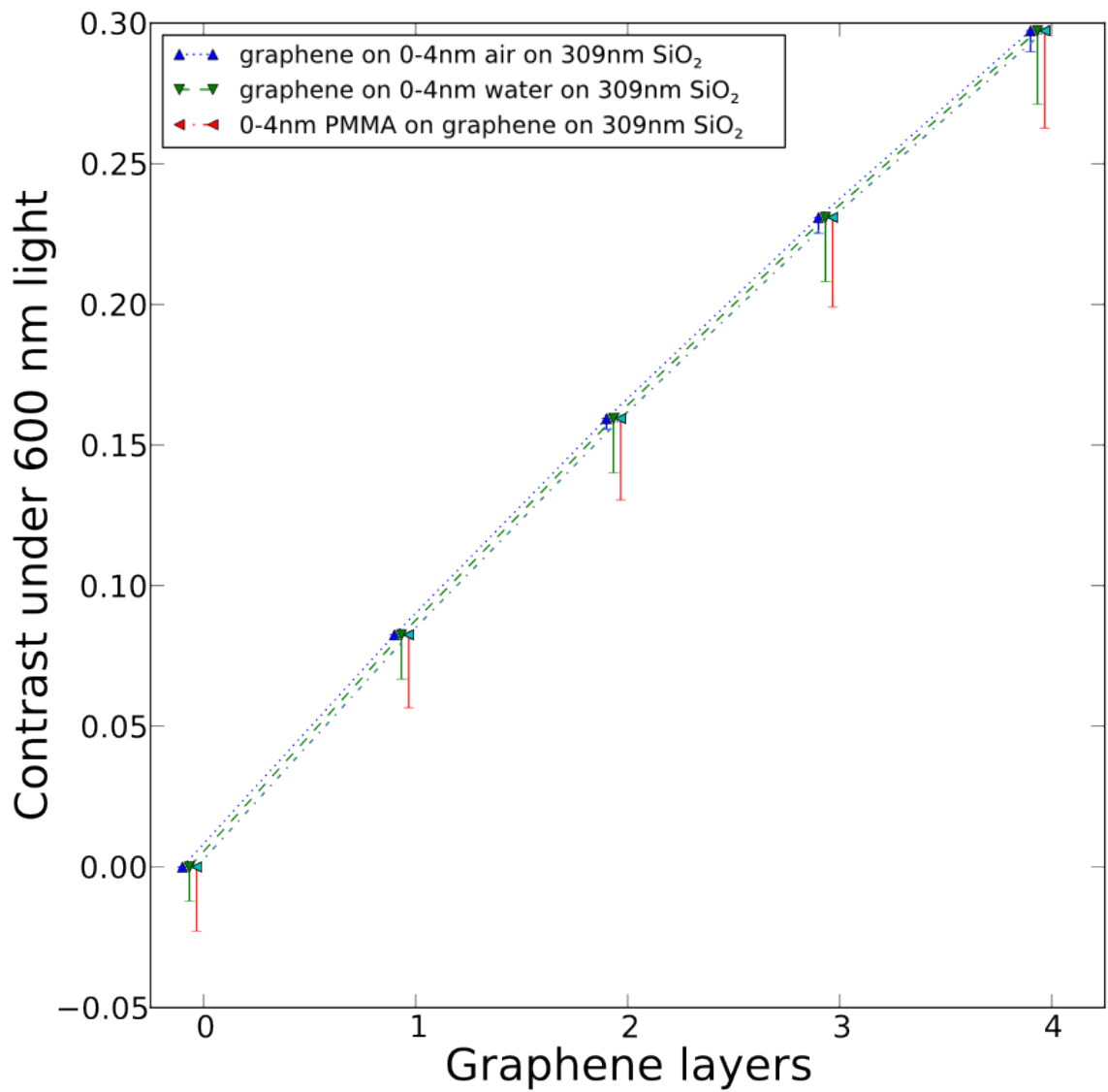

### Table S1: Raman data

Summary of Raman spectra of points 1-9 from Figure 1(a) in main text and Figure S1(a) using Lorentzian peak fits, along with spectra of the mechanically-exfoliated graphene bilayer flake shown in the insets of Figure 1(a) and Figure S1(a). All spectra show single *2D* peaks, except for the mechanically exfoliated bilayer, for which data on the *2D* peak is excluded. All peak center frequencies, $f_0$, and peak widths, $W$, have units of cm$^{-1}$, while relative peak intensity, $I/I_G$, is a unitless quantity.

| | Laser | *G* peak | | *D* peak | | | *2D* peak | | | *D'* peak | |
|---|---|---|---|---|---|---|---|---|---|---|---|
| Spot | $\lambda$ (nm) | $f_0$ (cm$^{-1}$) | $W$ | $f_0$ | $W$ | $I/I_G$ | $f_0$ | $W$ | $I/I_G$ | $f_0$ | $I/I_G$ |
| 1 | 514 | 1579.0 | 15.9 | 1341.9 | 26.1 | 0.21 | 2685.0 | 33.8 | 1.02 | 2456.4 | 0.05 |
| | 633 | 1582.5 | 13.6 | 1322.3 | 37.5 | 0.48 | 2652.0 | 32.5 | 1.78 | 2459.3 | 0.09 |
| 2 | 514 | 1580.0 | 21.5 | 1346.7 | 33.1 | 0.44 | 2693.1 | 43.0 | 0.71 | 2455.6 | 0.04 |
| | 633 | 1582.6 | 21.7 | 1327.3 | 38.8 | 1.20 | 2652.0 | 56.7 | 0.52 | 2467.1 | 0.06 |
| 3 | 514 | 1580.7 | 15.9 | 1342.8 | 25.1 | 0.25 | 2685.6 | 30.6 | 1.35 | 2459.4 | 0.06 |
| | 633 | 1586.3 | 10.6 | 1322.4 | 23.9 | 0.60 | 2642.9 | 31.0 | 1.24 | 2453.1 | 0.06 |
| 4 | 514 | 1582.4 | 21.0 | 1349.0 | 28.6 | 0.53 | 2697.5 | 40.2 | 0.73 | 2457.7 | 0.04 |
| | 633 | 1584.7 | 20.6 | 1329.4 | 31.2 | 1.40 | 2655.2 | 48.1 | 0.58 | 2469.8 | 0.06 |
| 5 | 514 | 1590.5 | 9.1 | 1348.8 | 16.6 | 0.22 | 2691.0 | 27.7 | 1.34 | 2466.4 | 0.13 |
| | 633 | 1589.0 | 11.2 | 1325.1 | 24.6 | 0.50 | 2646.6 | 31.3 | 1.25 | 2459.9 | 0.10 |
| 6 | 514 | 1591.7 | 14.5 | 1346.1 | 33.2 | 0.22 | 2692.8 | 34.4 | 1.37 | 2463.1 | 0.06 |
| | 633 | 1592.8 | 12.3 | 1324.8 | 24.9 | 0.47 | 2648.1 | 35.0 | 0.97 | 2475.3 | 0.04 |
| 7 | 514 | 1592.8 | 12.9 | 1347.8 | 23.8 | 0.19 | 2693.4 | 32.5 | 1.32 | 2462.6 | 0.06 |
| 8 | 514 | 1585.8 | 20.1 | 1355.3 | 40.9 | 0.12 | 2706.1 | 44.8 | 0.67 | 2461.5 | 0.04 |
| | 633 | 1584.0 | 20.3 | 1331.3 | 47.1 | 0.28 | 2658.9 | 50.9 | 0.46 | 2467.7 | 0.04 |
| 9 | 514 | 1586.8 | 18.1 | 1351.5 | 22.4 | 0.15 | 2699.0 | 38.3 | 0.84 | 2463.5 | 0.04 |
| | 633 | 1584.9 | 15.9 | 1330.5 | 36.0 | 0.18 | 2659.0 | 46.2 | 0.26 | 2466.0 | 0.02 |
| MEB | 514 | 1581.0 | 14.2 | 1352.8 | 15.2 | 0.01 | | | | 2457.5 | 0.05 |
| | 633 | 1581.2 | 12.4 | | | 0.00 | | | | 2471.5 | 0.04 |